\documentclass[onecolumn,showpacs]{revtex4}
\usepackage{graphicx}
\usepackage{dcolumn}
\usepackage{bm}

\input epsf

\begin{document}

\title{\Large  Interaction between Tachyon and Hessence (or Hantom) Dark Energies}

\author{\bf Surajit
Chattopadhyay$^1$\footnote{surajit$_{-}$2008@yahoo.co.in} and
~Ujjal Debnath$^2$\footnote{ujjaldebnath@yaghoo.com}}

\affiliation{ $^1$Department of Computer Application, Pailan
College of Management and Technology, Kolkata-104, India.\\
$^2$Department of Mathematics, Bengal engineering and Science
University, Howrah-103, India.}

\date{\today}

\begin{abstract}
In this paper, we have considered that the universe is filled with
tachyon, hessence (or hantom) dark energies. Subsequently we have
investigated the interactions between tachyon and hessence
(hantom) dark energies and calculated the potentials considering
the power law form of the scale factor. It has been revealed that
the tachyonic potential always decreases and hessence (or hantom)
potential increases with corresponding fields. Furthermore, we
have considered a correspondence between the hessence (or hantom)
dark energy density and new variable modified Chaplygin gas energy
density. From this, we have found the expressions of the arbitrary
positive constants $B_{0}$ and $C$ of new variable modified
Chaplygin gas.
\end{abstract}

\pacs{}

\maketitle

\section{\normalsize\bf{Introduction}}

Accelerated expansion of the universe is well documented in
literature [1]. This accelerated expansion is consistent with the
luminosity distance as a function of redshift of distant
Supernova, the structure formation and the cosmic microwave
background. A lot of recent observations have suggested that the
universe mainly consists of dark energy (73\%), dark matter
(23\%), baryon matter (4\%) and negligible radiation. To
accelerate the expansion, the equation-of-state parameter
$\omega=\frac{p}{\rho}$ of the dark energy must satisfy
$\omega<-\frac{1}{3}$, where $p$ and $\rho$ are its pressure and
energy density. The simplest candidate for its expansion is
cosmological constant $\Lambda$, for which the equation of state
is $\omega=-1$. However, there are several other evidences showing
that the dark energy might evolve from $\omega>-1$ in the past to
$\omega<-1$ today. The critical state $\omega=-1$ is crossed in
the intermediate redshift. Another possibility is quintessence [2]
which gives $-1\leq\omega\leq0$. However, the $k$-essence models
[3] and the phantom models [4] can get the state $\omega<-1$. As
the behaviour of $\omega$ crossing $-1$ can not be realized, more
complex models have suggested by several authors. The quintom
model, a hybrid of quintessence and phantom has been studied in
significant number of literature [5]. We consider the action

\begin{equation}
S=\int d^{4}x\sqrt{-g}\left(-\frac{{\cal R}}{16\pi G}+{\cal
L}_{DE}+{\cal L}_{m}\right)
\end{equation}

where, $g$ is the determinant of the metric $g_{\mu\nu}$, ${\cal
R}$ is the Ricci scalar, $\textit{L}_{DE}$ and $\textit{L}_{m}$
are the Lagrangian densities of dark energy and dark matter
respectively. The quintom dark energy has the Lagrangian density

\begin{equation}
{\cal
L}_{DE}=\frac{1}{2}[(\partial_{\mu}\phi_{1})^{2}+(\partial_{\mu}\phi_{2})^{2}]-V(\phi_{1},\phi_{2})
\end{equation}

where $\phi_{1}$ and $\phi_{2}$ are real scalar fields and play
the roles of quintessence and phantom respectively. In a spatially
flat FRW universe, under the assumption that $\phi_{1}$ and
$\phi_{2}$ are homogeneous, the effective equation of state is
given by

\begin{equation}
\omega=\frac{\dot{\phi_{1}}^{2}-\dot{\phi_{2}}^{2}-2V(\phi_{1},\phi_{2})}{\dot{\phi_{1}}^{2}-\dot{\phi_{2}}^{2}+2V(\phi_{1},\phi_{2})}
\end{equation}

It is obvious that for $\dot{\phi_{1}}^{2}>\dot{\phi_{2}}^{2}$ we
get $\omega>-1$ (i.e., quintessence model) and for
$\dot{\phi_{1}}^{2}<\dot{\phi_{2}}^{2}$ we get $\omega<-1$ (i.e.,
phantom model).\\

On the other hand there have been difficulties in obtaining
accelerated expansion from fundamental theories such as M/String
theory [6]. Much has been written and emphasized about the role of
the fundamental dilation field in the context of string cosmology.
But, not much emphasized is tachyon component [7]. It has been
recently shown by Sen [8, 9] that the decay of an unstable D-brane
produces pressure-less gas with finite energy density that
resembles classical dust. The cosmological effects of the tachyon
rolling down to its ground state have been discussed by Gibbons
[10]. Rolling tachyon matter associated with unstable D-branes has
an interesting equation of state which smoothly interpolates
between $-1$ and 0. As the Tachyon field rolls down the hill, the
universe experiences accelerated expansion and at a particular
epoch the scale factor passes through the point of inflection
marking the end of inflation [6]. The tachyonic matter might
provide an explanation for inflation at the early epochs and could
contribute to some new form of cosmological dark matter at late
times [11]. Inflation under tachyonic field has also been
discussed in ref. [7, 12, 13]. Sami et al [14] have discussed the
cosmological
prospects of rolling tachyon with exponential potential.\\

The action for the homogeneous tachyon condensate of string theory
in a gravitational background is given by,
\begin{equation}
S=\int {\sqrt{-g}~ d ^{4} x \left[\frac{\cal R}{16 \pi G}+{\cal
L}\right]}
\end{equation}
where $\cal L$ is the Lagrangian density given by,
\begin{equation}
{\cal {L}}=-V(\phi)\sqrt{1+g^{\mu \nu}~\partial{_{\mu}}\phi
\partial{_{\nu}} \phi}
\end{equation}
where $\phi$ is the tachyonic field, $V(\phi)$ is the tachyonic
field potential and $\cal R$ is the Ricci Scalar.\\

 The Chaplygin gas is characterized by an
exotic equation of state $p=-\frac{B}{\rho}$, where $B$ is a
positive constant. Role of Chaplygin gas in the accelerated
universe has been studied by several authors [15]. The above
mentioned equation of state has been modified to
$p=-\frac{B}{\rho^{\alpha}}$ with $0\leq\alpha\leq 1$. This is
called generalized Chaplygin gas [16]. This equation has been
further modified to $p=A\rho-\frac{B}{\rho^{\alpha}}$ with
$0\leq\alpha\leq 1$. This is called modified Chaplygin gas [17].
This equation of state shows radiation era at one extreme and
$\Lambda CDM$ model at the other extreme. Debnath [18] introduced
a variable modified Chaplygin gas with $B$ as a function of the
scale factor $a$ and the
equation of state is $p=A\rho-\frac{B(a)}{\rho^{\alpha}}$.\\
Although the two dark components are usually studied under the
assumption that there is no interaction between them, one can not
exclude such a possibility. In fact, researches show that a
presumed interaction may help alleviate the coincidence problem
[19]. Some models that allow interaction between the scalar field
and the matter field have been proposed as a solution to the
cosmic coincidence problem [20]. Models based on dark energy
interacting with dark matter have been widely investigated by
several authors. These models yield stable scaling solution of the
FRW equations at late times of the evolving universe. Interacting
Chaplygin gas allows the universe to cross the phantom divide,
which is not possible by the pure Chaplygin gas [20]. There is a
report that this interaction is physically observed in the Abell
cluster A586, which in fact supports the GCG cosmological model
and apparently rules out the $\Lambda$CDM model [21]. Herrera et
al (2004) [22] considered interacting mixture of cold dark matter
and a tachyonic field. In this study they assumed that both
components -the tachyon field and the cold dark matter- do not
conserve separately but that they interact through a term $Q$ and
found exact solutions leading to power law accelerated expansion
for a homogeneous, isotropic and spatially flat universe,
dominated by an interacting mixture of cold dark matter and a
tachyonic field. Setare et al (2009) [23] considered an
interaction between the tachyonic field and the barotropic fluid
and obtained the pressure and energy densities by choosing the
interaction term $Q$ as proportional to the density of the
barotropic fluid and the Hubble parameter. Similar choice of
interaction term $Q$ was adopted in Amendola et al (2007) [24],
who considered dark matter-dark energy interaction and discussed
its consequences on cosmological parameters derived from SNIa
data. In a recent paper, Sheykhi (2010) [25] demonstrated that the
interacting agegraphic evolution of the universe can be described
completely by a single tachyon scalar field and thus reconstructed
the potential as well as the dynamics of the tachyon field
according to the evolutionary behavior of interacting agegraphic
dark energy. Wei and Cai (2005) [26] considered an interaction
between hessence and and background perfect fluid through an
interaction term described earlier. In the context of the
literature surveyed, we decided to consider a two-component model,
where instead of dark energy and dark matter components, we
decided to consider two candidates of dark energy (tachyon and
hessence) in an interacting situation. We reconstructed the scalar
fields and potentials of both of the candidates under interaction.
Furthermore, we studied a correspondence hessence and new variable
modified Chaplygin gas.
\\
The organization of the paper is as follows: Section II describes
the Lagrangian form of hessence and hantom dark energies. In
section III, we have investigated the interaction between tachyon
and hessence (or hantom) dark energies. In section IV, we have
shown hessence (or hantom) as new variable modified Chaplygin gas
and found the expressions of the arbitrary positive constants of
new variable modified Chaplygin gas. Finally, the paper ends with
a short discussion in section V.
\\

\section{\normalsize\bf{Hessence and Hantom Dark Energies}}

The non-canonical complex scalar field has been given the name
``hessence" [27]. The Lagrangian density of hessence is given by
\begin{equation}
{\cal L}_{he}=\frac{1}{4}[(\partial_{\mu}
\phi_{1})^{2}+(\partial_{\mu} \phi_{2})^{2}]-
V(\phi_{1}^{2}-\phi_{2}^{2})
\end{equation}
 Now $\phi_{1}^{2}-\phi_{2}^{2}$ = constant is a hyperbola on the
 $\phi_{1}$ versus $\phi_{2}$ plane and using the relations
 between angular functions and hyperbolic functions we can define
 the transformations given by

\begin{equation}
\phi_{1}\longrightarrow
\phi_{1}cosh(i\alpha)-\phi_{2}sinh(i\alpha)\quad,\quad\quad
\phi_{2}\longrightarrow-\phi_{1}sinh(i\alpha)+\phi_{2}cosh(i\alpha)
\end{equation}

Now consider two new variables $\phi$ and $\theta$ to describe the
hessence as
\begin{equation}
\phi_{1}= \phi cosh \theta\quad,\quad\quad \phi_{2}=\phi sinh
\theta
\end{equation}
which are defined by

\begin{equation}
\phi^{2}=\phi_{1}^{2}-\phi_{2}^{2} \quad,\quad\quad coth \theta
=\frac{\phi_{1}}{\phi_{2}}
\end{equation}
then the transformation equation (7) is equivalent to
\begin{equation}
 \phi \longrightarrow \phi \quad\quad\quad and
\quad\quad\quad\theta \longrightarrow \theta - i \alpha
\end{equation}
\ which means an internal imaginary motion. Thus formulation of
$(\phi,\theta)$ is to describe the non-canonical complex scalar
field. Now we can write the Lagrangian density of the hessence in
the form:

\begin{equation}
{\cal L}_{he}=\frac{1}{2}[(\partial_{\mu}
\phi)^{2}-\phi^{2}(\partial_{\mu} \theta)^{2}]- V(\phi)
\end{equation}

The Lagrangian density of hantom is given by
\begin{equation}
{\cal L}_{he}=\frac{1}{4}[(\partial_{\mu}
\phi_{1})^{2}+(\partial_{\mu} \phi_{2})^{2}]-
V(\phi_{2}^{2}-\phi_{1}^{2})
\end{equation}

Again consider two variables $\phi$ and $\theta$ to describe the
hantom as
\begin{equation}
\phi_{1}= \phi sinh \theta\quad,\quad\quad \phi_{2}=\phi cosh
\theta
\end{equation}
which are defined by

\begin{equation}
\phi^{2}=\phi_{2}^{2}-\phi_{1}^{2} \quad,\quad\quad coth \theta
=\frac{\phi_{2}}{\phi_{1}}
\end{equation}

Now we can write the Lagrangian density of the hantom in the form

\begin{equation}
{\cal L}_{he}=-\frac{1}{2}[(\partial_{\mu}
\phi)^{2}-\phi^{2}(\partial_{\mu} \theta)^{2}]- V(\phi)
\end{equation}

\section{\normalsize\bf{Interaction between Tachyon and Hessence (or Hantom) Dark Energies}}

Considering the spatially flat FRW universe with metric

\begin{equation}
ds^{2}= dt^{2}-a^{2}(t)\left[dr^{2}+
r^{2}(d\theta^{2}+\sin^{2}\theta  d\phi^{2})\right]
\end{equation}

where $a(t)$ is the expansion scalar or the scale factor. From
(11) and (15) we have the pressure and energy densities of
hessence (or hantom) as

\begin{equation}
p_{h}=\frac{\epsilon}{2}(\dot{\phi}^{2}-\phi^{2}\dot{\theta}^{2})-V(\phi)
\end{equation}
and
\begin{equation}
\rho_{h}=\frac{\epsilon}{2}(\dot{\phi}^{2}-\phi^{2}\dot{\theta}^{2})+V(\phi)
\end{equation}

where $\dot{\theta}=\frac{Q}{a^{3}\phi^{2}}$, $Q$ is the total
conserved energy = constant. The symbol $\epsilon$ is used to
represent hessence ($\epsilon=1$) and hantom ($\epsilon=-1$) using
one equation for pressure and equation for density. To consider
the interaction between the hessence (or hantom) and tachyon, we
denote $V(\phi)$ of tachyon as $V_{1}$ and that of hessence (or
hantom) as $V_{2}$. The fields of tachyon and hessence (or hantom)
are denoted as $\phi_{1}$ and $\phi_{2}$ respectively. To denote
pressure and energy densities of tachyon we use 1 in subscript and
to denote
hessence (or hantom) we use 2 in subscript.\\

Thus the pressure and energy densities of tachyonic field are
given by

\begin{equation}
p_{1}=-V_{1}(\phi_{1})\sqrt{1-\dot{\phi}_{1}^{2}}
\end{equation}
and
\begin{equation}
\rho_{1}=\frac{V_{1}(\phi_{1})}{\sqrt{1-\dot{\phi_{1}}^{2}}}
\end{equation}

The pressure and energy densities of hessence (or hantom) are

\begin{equation}
p_{2}=\epsilon\left(\frac{1}{2}\dot{\phi_{2}}^{2}-\frac{Q^{2}}{2a^{6}\phi_{2}^{2}}\right)-V_{2}(\phi_{2})
\end{equation}
and
\begin{equation}
\rho_{2}=\epsilon\left(\frac{1}{2}\dot{\phi_{2}}^{2}-\frac{Q^{2}}{2a^{6}\phi_{2}^{2}}\right)+V_{2}(\phi_{2})
\end{equation}

Introducing the interaction parameter $\delta$, the equations for
continuity becomes

\begin{equation}
\dot{\rho_{1}}+3H(\rho_{1}+p_{1})=3H\delta\rho_{2}
\end{equation}

and

\begin{equation}
\dot{\rho_{2}}+3H(\rho_{2}+p_{2})=-3H\delta\rho_{2}
\end{equation}

From the equation (24) we get

\begin{equation}
\epsilon\left[\ddot{\phi_{2}}+3H\left(1+\frac{\delta}{2}\right)\dot{\phi_{2}}\right]+\left[\frac{\dot{V_{2}}}{\dot{\phi_{2}}}
+3H\delta\frac{V_{2}}{\dot{\phi_{2}}}+\epsilon\left(\frac{Q^{2}}{{\phi_{2}}^{3}a^{6}}-\frac{3H\delta
Q^{2}}{2a^{6}\phi_{2}^{2}\dot{\phi_{2}}}\right)\right]=0
\end{equation}

Let us assume,

\begin{equation}
\frac{dV_{3}}{d\phi_{2}}=\frac{\dot{V_{2}}}{\dot{\phi}_{2}}+3H\delta\frac{V_{2}}{\dot{\phi_{2}}}+\epsilon\left(\frac{Q^{2}}{\phi_{2}^{3}a^{6}}
-\frac{3H\delta Q^{2}}{2a^{6}\phi_{2}^{2}\dot{\phi_{2}}}\right)
\end{equation}

where, $V_{3}$ is a function of $\phi_{2}$. Thus, from equations
(25) and (26), we get

\begin{equation}
\epsilon\ddot{\phi_{2}}+3H\epsilon\left(1+\frac{\delta}{2}\right)\dot{\phi_{2}}+\frac{dV_{3}}{d\phi_{2}}=0
\end{equation}

For simplicity, let us choose $V_{3}=n\dot{\phi_{2}}^{2}$ and
$a=a_{0}t^{m}$, we get

\begin{equation}
{\phi_{2}}=\frac{C_{0}}{1+\alpha}a_{0}^{\frac{1}{m}}t^{1+\alpha}
\end{equation}

where $a_{0}$ and $C_{0}$ are positive constants,
$\alpha=-\frac{3m\epsilon}{2}(\frac{2+\delta}{2n+1})$ and for this
$\alpha$ we get

\begin{equation}
V_{3}=n C_{0}^{2}a_{0}^{\frac{2\alpha}{m}}t^{2\alpha}
\end{equation}

\begin{figure}
\includegraphics[height=1.8in]{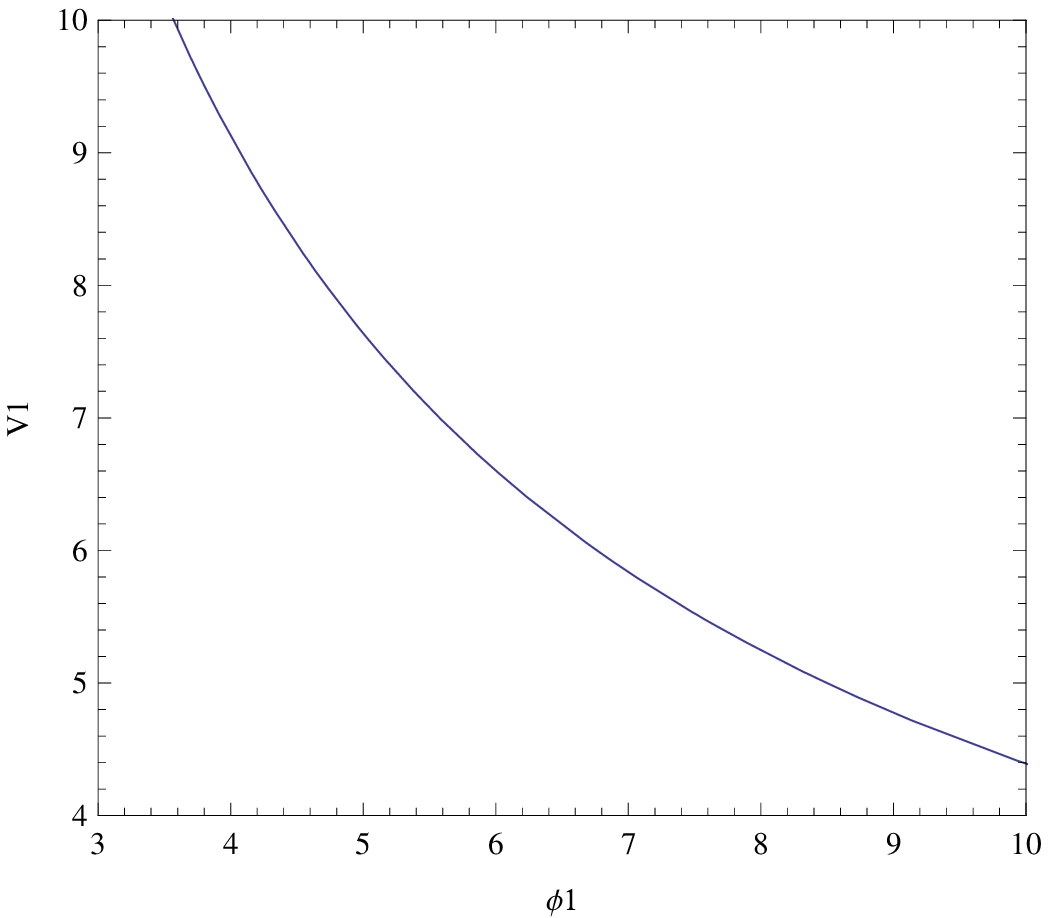}~~~~
\includegraphics[height=1.8in]{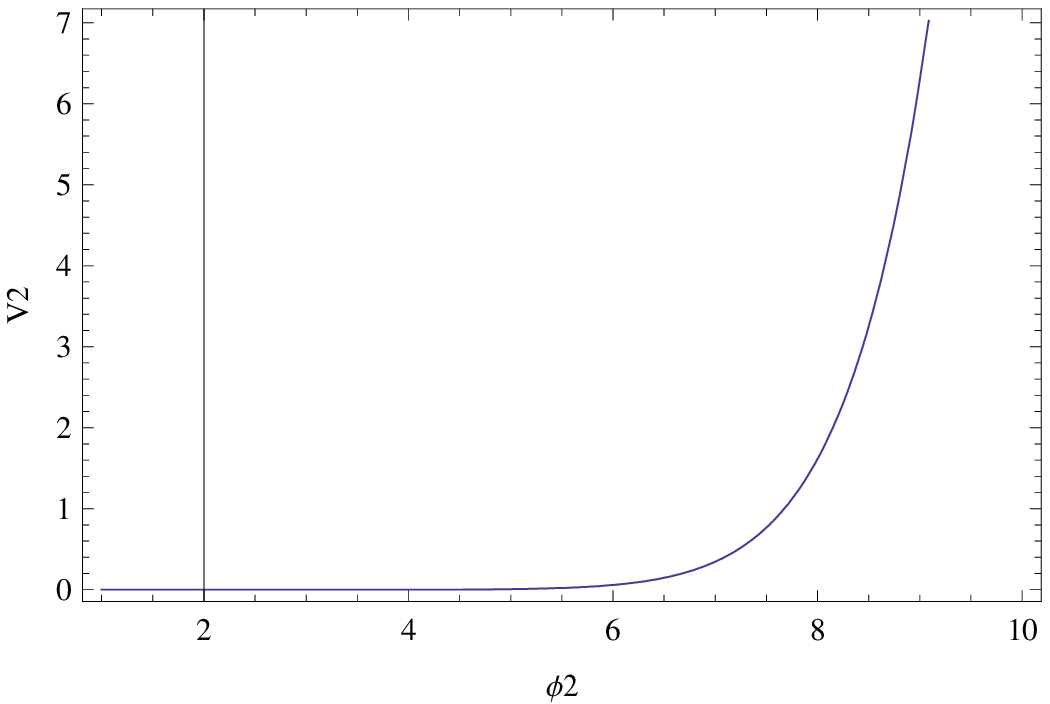}\\
\vspace{1mm}
~~~~~Fig.1~~~~~~~~~~~~~~~~~~~~~~~~~~~~~~~~~~~~~~~~~~~~~~~~~Fig.2\\
\vspace{4mm}
\includegraphics[height=1.8in]{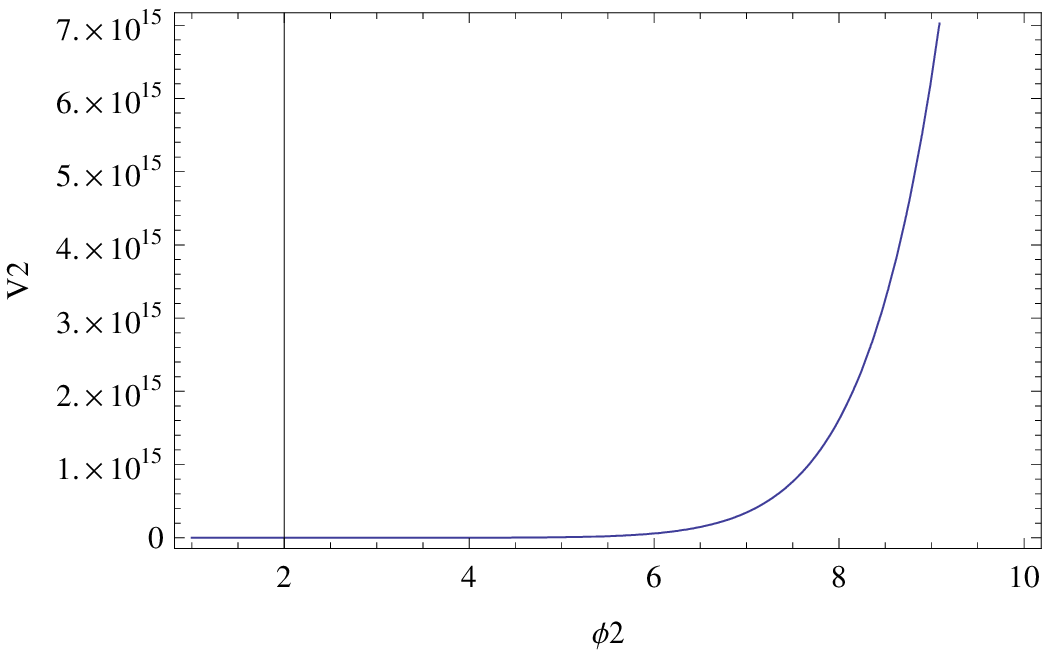}\\
\vspace{1mm}~~~~~~~~~~~~~~~Fig.3

\vspace{6mm}

Figs. 1, 2 and 3 represent the variations of potentials against
tachyon, hessence ($\epsilon=+1$) and hantom ($\epsilon=-1$)
fields respectively.

 \vspace{7mm}

\end{figure}

Therefore, from equations (26), (28) and (29), it can be obtained
that the expression for hessence (or hantom) potential $V_{2}$ as

\begin{eqnarray*}
V_{2}=C_{1}\left(\frac{(1+\alpha)a_{0}^{-\frac{\alpha}{m}}\phi_{2}}{C_{0}}
\right)^{-\frac{3m\delta}{1+\alpha}}+\frac{2\alpha
C_{0}^{2}a_{0}^{6+\frac{2\alpha}{m(1+\alpha)}}}{2\alpha+3m\delta}\left(\frac{(1+\alpha)\phi_{2}}{C_{0}}
\right)^{\frac{2\alpha}{1+\alpha}}+~~~~~~~~~~~~~~~~~~~~~~~~~~
\end{eqnarray*}
\begin{equation}
\frac{4na_{0}^{6}C_{0}^{4}Q^{2}\epsilon\left\{2(1+\alpha)-3m\delta
a_{0}^{-\frac{2\alpha}{m}}\right\}}{2(1+\alpha)-3m(-2+\delta)}\left(\frac{(1+\alpha)\phi_{2}a_{0}^{-\frac{\alpha}{m}}}{C_{0}}
\right)^{-\frac{2(1+\alpha+3m)}{1+\alpha}}
\end{equation}

where $C_{1}$ is arbitrary integration constant. Now from equation
(23), we get

\begin{equation}
P=\frac{2-3\dot{\phi_{1}}^{2}}{\sqrt{1-\dot{\phi_{1}}^{2}}}
\end{equation}

where,

\begin{eqnarray*}
P=-\frac{6m(m-1)}{t^{2}}-2C_{1}t^{-3m\delta}+2a_{0}^{\frac{2\alpha}{m}}t^{2\alpha}\left(C_{0}^{2}-\frac{2\alpha
C_{0}^{2}a_{0}^{6}}{2\alpha+3m\delta}\right)-~~~~~~~~~~~~~~~~~~~~~~~~~~~~~~~
\end{eqnarray*}
\begin{equation}
2(1+\alpha)^{2}Qt^{-2(1+\alpha+3m)}\left[\frac{a_{0}^{-6-\frac{2\alpha}{m}}}{C_{0}^{2}}+
\frac{4na_{0}^{6}C_{0}^{4}Q\epsilon\left\{2(1+\alpha)-3m\delta
a_{0}^{-\frac{2\alpha}{m}}\right\}}{2(1+\alpha)-3m(-2+\delta)}\right]
\end{equation}

Furthermore, it can be obtained that the expressions for tachyonic
potential $V_{1}$ and tachyonic field $\phi_{1}$ as

\begin{equation}
V_{1}=C_{1}t^{-3m\delta}+\frac{\alpha
t^{2\alpha-1}}{2C_{0}^{2}a_{0}^{6}(-1+2\alpha+3m\delta)}
+\frac{2C_{0}^{2}a_{0}^{6}(1+\alpha)^{2}Q^{2}t^{-2(1+\alpha+3m)}\{\epsilon(1+\alpha+3m)+\alpha(1+2n)
\} }{2(1+\alpha)-3m(-2+\delta)}
\end{equation}
and
\begin{equation}
\phi_{1}=\int\sqrt{\frac{2}{3}-\frac{P^{2}}{18}-\frac{P}{18}\sqrt{12+P^{2}}}~dt
\end{equation}

From above expressions, we see that expression of $\phi_{1}$ is
very complicated, so $V_{1}$ can not be expressed in terms of
$\phi_{1}$ explicitly. So some numerical investigations are needed
to see the nature of tachyonic potential. From figure 1, it is
discerned that the tachyonic potential is decreasing with the
field and figure 2, 3 show that the hessence and hantom potentials
are increasing with the corresponding fields for
$m=1,n=2,a_{0}=1$.

\section{\normalsize\bf{Hessence (or Hantom) as New Variable Modified Chaplygin Gas}}

In the present section, we consider hessence (or hantom) as  gas
as new variable modified Chaplygin gas. The endeavor is to
establish a correspondence between the hessence (or hantom) and
the variable modified Chaplygin gas model. Assuming
$B(a)=B_{0}a^{-n}$ in the equation of state of the variable
modified Chaplygin gas with $B_{0}>0$ and $n$ as positive constant
we get the solution for $\rho$ as

\begin{equation}
\rho_{\Lambda}=\left[\frac{3(1+\alpha)B_{0}}{3(1+\alpha)(1+A)-n}\frac{1}{a^{n}}+
\frac{C}{a^{3(1+A)(1+\alpha)}}\right]^{\frac{1}{1+\alpha}}
\end{equation}

Taking derivatives of both sides with respect to cosmic time we
get

\begin{equation}
\dot{\rho}_{\Lambda}=3H\left[Ca^{-(1+A)(1+\alpha)}+\frac{3(1+\alpha)a^{-n}B_{0}}{3(1+A)(1+\alpha)-n}\right]^{-\frac{\alpha}{1+\alpha}}
\left[-(1+A)Ca^{-3(1+A)(1+\alpha)}-\frac{na^{-n}B_{0}}{3(1+A)(1+\alpha)-n}\right]
\end{equation}

Using the equation of density (22) for hessence (or hantom) (write
$\rho_{\Lambda}$ instead of $\rho_{2}$) we get

\begin{equation}
\dot{\rho}_{\Lambda}=-32^{\alpha}H\left[(1+A)Ca^{-3(1+A)(1+\alpha)}+\frac{na^{-n}B_{0}}{3(1+A)(1+\alpha)-n}\right]\left[-2V_{2}
+\epsilon\left(-\frac{Q^{2}}{a^{6}\phi_{2}^{2}}+\dot{\phi}_{2}^{2}\right)\right]^{-\alpha}
\end{equation}

and

\begin{equation}
p_{\Lambda}=2^{\alpha}a^{-n}\left[ACa^{n-3(1+A)(1+\alpha)}
+\frac{(n-3(1+\alpha)B_{0})}{3(1+A)(1+\alpha)-n}\right]\left[-2V_{2}+\epsilon\left(-\frac{Q}{a^{6}\phi_{2}^{2}}+\dot{\phi_{2}}^{2}\right)\right]^{-\alpha}
\end{equation}

From the equations of continuity the effective equation of state
is

\begin{equation}
\omega_{\Lambda}^{eff}=A-\frac{B_{0}a^{-n}}{\rho_{\Lambda}^{1+\alpha}}+\frac{\delta}{3H}
\end{equation}

 Following reference [28] in the above equation we use
$\delta=\frac{3b^{2}H(1+\Omega_{k})}{\Omega_{\Lambda}}$ to get

\begin{equation}
\omega_{\Lambda}^{eff}=A-\frac{B_{0}a^{-n}}{(3M_{p}^{2}H^{2}\Omega_{\Lambda})^{1+\alpha}}+\frac{b^{2}(1+\Omega_{k})}{\Omega_{\Lambda}}
\end{equation}

where,

\begin{equation}
\Omega_{\Lambda}=\frac{\rho_{\Lambda}}{3M_{p}^{2}H^{2}}~,~
\Omega_k=\frac{k}{a^{2}H^{2}}
\end{equation}

In non-flat universe, our choice for hessence (or hantom) energy
density is

\begin{equation}
\rho_{\Lambda}=-V_{2}+\frac{\epsilon}{2}\left(-\frac{Q^{2}}{a^{6}\phi_{2}^{2}}+\dot{\phi}_{2}^{2}\right)
\end{equation}

we get (using the continuity equations (21) and (22)) the values
of arbitrary constants $B_{0}$ and $C$ as

\begin{equation}
B_{0}=\frac{a^{n}\{3(1+A)(1+\alpha)-n\}}{3(1+\alpha)}\left[\left\{-V_{2}+\frac{\epsilon}{2}\left(-\frac{Q^{2}}{a^{6}\phi_{2}^{2}}+
\dot{\phi}_{2}^{2}\right)\right\}^{1+\alpha}
-\frac{C}{a^{3(1+\alpha)(1+A)}}\right]
\end{equation}

\begin{eqnarray*}
C=\frac{3(1+\alpha)a^{3(1+A)(1+\alpha)}}{3(1+A)(1+\alpha)-n}\left[(3M_{p}^{2}H^{2}\Omega_{\Lambda})^{1+\alpha}
\left(-\frac{1}{3}-A-\frac{b^{2}(1+\Omega_{k})}{\Omega_{\Lambda}}
-\frac{2\sqrt{\Omega_{\Lambda}-c^{2}\Omega_{k}}}{3c}\right)\right.
\end{eqnarray*}
\begin{equation}
 \left.+\frac{\{3(1+A)(1+\alpha)-n\}}{3(1+\alpha)}
\left\{-V_{2}+\frac{\epsilon}{2}\left(-\frac{Q^{2}}{a^{6}\phi_{2}^{2}}+
\dot{\phi}_{2}^{2}\right)\right\}^{1+\alpha}\right]
\end{equation}

\section{\normalsize\bf{Conclusions}}

In this paper, we have considered that the universe is filled with
tachyon, hessence (or hantom) dark energies. Subsequently we have
investigated the interactions between tachyon and hessence (or
hantom) dark energies and calculated the potentials considering
the power law form of the scale factor. It has been revealed that
the tachyonic potential $V_{1}$ always decreases with $\phi_{1}$
and hessence (or hantom) potential $V_{2}$ increases with
corresponding field $\phi_{2}$. Furthermore, we have considered a
correspondence between the hessence (or hantom) dark energy
density and new variable modified Chaplygin gas energy density.
From this, we have found the expressions of the arbitrary positive
constants $B_{0}$ and $C$ of new variable modified
Chaplygin gas. \\

{\bf Acknowledgement:}\\

The authors are thankful to IUCAA, Pune, India for warm
hospitality where part of the work was carried out. Also UD is
thankful to UGC, Govt. of India for providing research project grant (No. 32-157/2006(SR)).\\

{\bf References:}\\
\\
$[1]$  N. A. Bachall, J. P. Ostriker, S. Perlmutter and P. J.
Steinhardt, {\it Science} {\bf 284} 1481 (1999); S. J. Perlmutter
et al, {\it Astrophys. J.} {\bf 517} 565 (1999); V. Sahni and A.
A. Starobinsky, {\it Int. J. Mod. Phys. A} {\bf 9} 373 (2000); P.
J. E. Peebles and B. Ratra, {\it Rev. Mod. Phys.} {\bf 75} 559
(2003); T. Padmanabhan, {\it Phys. Rept.} {\bf 380} 235 (2003); E.
J. Copeland, M. Sami, S. Tsujikawa, {\it Int. J. Mod. Phys. D}
{\bf  15} 1753 (2006).\\
$[2]$ C. Wetterich, {\it Nucl. Phys. B} {\bf 302} 668 (1988).\\
$[3]$ C. Armendariz-Picon, V.F. Mukhanov and P. J. Steinhardt,
    {\it Phys. Rev. D} {\bf 63} 103510 (2001).\\
$[4]$ R. R. Caldwell, {\it Phys. Lett. B} {\bf 545} 23 (2002);
S. Nojiri, S. D. Odintsov, {\it Phys. Rev. D} {\bf 72} 023003 (2005).\\
$[5]$ B. Feng, X. L. Wang and X. M. Zhang, {\it Phys. Lett. B },
    {\bf 607} 35 (2005); X. Zhang, {\it Commun. Theor. Phys.} {\bf
    44} 762 (2005).\\
$[6]$ M. Sami, {\it Mod. Phys. Lett. A} {\bf 18} 691 (2003).\\
$[7]$ A. Feinstein, {\it Phys. Rev. D} {\bf 66} 063511 (2002).\\
$[8]$ A. Sen, {\it JHEP} {\bf 0204} 048 (2002).\\
$[9]$ A. Sen, {\it JHEP} {\bf 0207} 065 (2002).\\
$[10]$ G. W. Gibbons, {\it Phys. Lett. B} {\bf 537} 1 (2002).\\
$[11]$ M. Sami, P. Chingangbam and T. Qureshi, {\it Phys. Rev. D} {\bf 66} 043530 (2002).\\
$[12]$ M. Fairbairn and M. H. G. Tytgat, {\it Phys. Lett. B} {\bf 546} 1 (2002).\\
$[13]$ T. Padmanabhan, {\it Phys. Rev. D} {\bf 66} 021301 (2002).\\
$[14]$ M. Sami, P. Chingangbam and T. Qureshi, {\it Pramana} {\bf 62} 765 (2004).\\
$[15]$ A. Kamenshchik, U. Moschella and V. Pasquier, {\it Phys.
Lett. B} {\bf 511} 265 (2001); V. Gorini, A. Kamenshchik, U.
Moschella and V. Pasquier, {\it gr-qc}/0403062.\\
$[16]$ V. Gorini, A. Kamenshchik and U. Moschella, {\it Phys. Rev.
D} {\bf 67} 063509 (2003); U. Alam, V. Sahni , T. D. Saini and
A.A. Starobinsky, {\it Mon. Not. Roy. Astron. Soc.} {\bf 344},
1057 (2003); M. C. Bento, O. Bertolami and A. A. Sen, {\it Phys.
Rev. D} {66} 043507 (2002).\\
$[17]$ H. B. Benaoum, {\it hep-th}/0205140; U. Debnath, A.
Banerjee and S. Chakraborty, {\it Class.
Quantum Grav.} {\bf 21} 5609 (2004).\\
$[18]$ U. Debnath, {\it Astrophys. Space Sc.} {\bf 312} 295 (2007).\\
$[19]$ X. Fu, H. Yu, P. Wu, {\it Phys. Rev. D} {\bf 78} 063001 (2008).\\
$[20]$ F. P. Zen et al, {\it Eur. Phys J. C} {\bf 63} 477 (2009).\\
$[21]$ M. Jamil and M. A. Rashid, {\it Eur. Phys J. C} {\bf 58}
111
(2008).\\
$[22]$ R. Herrera, D. Pavón, W. Zimdahl, {\it General Relativity
and Gravitation} {\bf 36} 2161 (2004).\\
$[23]$ M. R. Setare, J. Sadeghi, A. R. Amani, {\it Phys. Lett. B}
{\bf 673} 241 (2009).\\
$[24]$ L. Amendola, G. C. Campos, R. Rosenfeld, {\it Phys. Rev. D}
{\bf 75} 083506 (2007).\\
$[25]$ A. Sheykhi, {\it Phys. Lett. B} {\bf 682} 329 (2010).\\
$[26]$ H. Wei, R.-G. Cai, {\it Phys. Rev. D} {\bf 72} 123507
(2005).\\
$[27]$ H. Wei, R.G. Cai and D.F. Zeng, {\it Class. Quantum Grav.}
    {\bf 22} 3189 (2005).\\
$[28]$ M. R. Setare, {\it Europhys. J. C} {\bf 52} 689 (2007).\\

\end{document}